# Optimal performance of heterogeneous networks based on the bit rate

**Déthié Dione**[1] · **Salimata Gueye Diagne** [1] · **Bakary Koné** [1] · **Youssou Gningue**[2]

**Abstract** Networks LTE(4G) and Wi-Fi complementarity establish a heterogeneous system of wireless and mobile networks. We study and analyze the optimal performances of this heterogeneous system based on the bit rate, the blocking probability and user connection loss. Random Waypoint is the user mobility model. User's provided mobile terminal equipped with multiple accesses interfaces. We have developed a Markov chain to estimate the performances obtained from the heterogeneous networks system, which allowed us to propose an average bit rate value in a sub-zone of this system then the average blocking probability user connection in this zone. We have also proposed a sensitivity factor of maximal decrease of these selection network parameters. This factor informs about the heterogeneous networks congestion and dis-congestion system.

## Introduction

The integration of wireless and mobile networks such as longterm evolution (LTE) and Wi-Fi is nowadays a necessity for the satisfaction of user request which is stronger and stronger. The global services and user mobility makes this task difficult. However these networks are provided with characteristics to support services and user mobility.

✉ Déthié Dione
dethiedione79@gmail.com

1 Department of Mathematic and Computer Science, University of Cheikh Anta Diop, Dakar, Senegal

2 University of Laurentia, Sudbury, Canada

The Random waypoint (RWP) is model chosen for users mobility in this heterogeneous networks system. The RWP model corresponds to the ideal behavior of a user in an urban area because according to the model of mobility RWP, every user chooses randomly a place of destination and goes to this one in a constant velocity. The user movement starting up from a point to a destination is named ”one movement epoch”. The users velocity in every epoch is random variable and is chosen from a uniform distribution of the velocity [0, *V*max], where *V*max is the eligible maximal velocity for user. In the RWP, the user can wait for a period of time called ”time of reflection” before his departure for another point. The browsed path of a user is independent from its previous path and from other users path. Thus, at the end of every time of movement, the user stops a duration of the time and then chooses another destination place and, possibly, new velocity, and moves towards this destination in a constant velocity, and so on. RWP is one of the mobility models widely used in the performances analysis of the wireless and mobile networks. It represents well the individual movements which include the stop, the starting up and the other actions bound to the individual movements in cities.

So new methods for saving, transmission and sharing of bandwidth are imperative. Among these we can mention the selection technique of best network based on the bit rate which the selection parameters of which are the blocking probability and connections losses.

We noted in the literature the most used selection strategies of network. In their works [10], we were able to analyze the signal power received (RSNS) and then the available bandwidth (TBNS) of a heterogeneous networks system. They emphasized the parameters of this system such as the blocking probability and connections losses but they did not take into account the interference in the selection techniques that they developed which made less successful results obtained from the blocking probability and connections losses. Besides the authors [1, 2, 12, 13] considered the interference in the selection strategies they adopted which is the one based on the SINR, which allowed them to decrease the probability connections losses during a vertical handover. But through their studies, they did not approach the parameters connections blocking. On the other hand, these are analyzed by [7] who in their works obtained results more satisfactory than those who are preceded them. Besides, [11] took into account the users mobility(terminal-controlled mobility management) and other aspects such as the cost, the battery life cycle and the handover frequency.

However, through their studies, none of these authors took into account the constraint related to the block error rate (BLER) in a sub-cell given by the cluster. The contribution in our works bases on our model users connected and disconnected. When he is connected, the system quits a well-defined state and moves to an other state before returning to this first one in a given time interval. Besides, unlike the other works, we took the bit rate as an important selection technique of the best network which consists in choosing the biggest bit rate value.

The paper is organized as follows: “Model of heterogeneous networks system” section introduces the model of the studied heterogeneous networks where all the parameters of the system are defined. The selection method algorithm based on the bit rate is established at the level of “Selection method” section. In “User mobility model” section, we have developed the users mobility model, it is Random Waypoint (RWP) which we consider more adequate to the individual users movements. The average access demand rate for a service is given in “Average new access demand rate $\lambda_{C_i}^{C(k)}$ for a service” section. The average rate demand of vertical and horizontal handover is calculated in “Average demand rate of handovers” section. In “Modeling approach based on a Markov chain” section, we have used a Markov chain to analyze the number busy bandwidth units. The system studied performances such as the bit rate and the blocking probability and connection losses are estimated in “Evaluations of optimal performances” section. The results obtained are simulated in “Numerical tests” section. We finished our study in “Conclusion” section by conclusion and future work.

## Model of heterogeneous networks system

The model of heterogeneous networks system which we study is represented by Fig. 1.

Indeed, we have an hexagonal area of service $C_1$ covered entirely by the mobile network LTE (4G). In this service area are present several homogeneous circular sub-cells $(C_j)_{2\leq j\leq m}$ of radius $r_i$ among which each is also covered by a Wi-Fi wireless. So both mobile network LTE and Wi-Fi wireless overlap in cells $C_i$ and the Wi-Fi wireless are separated between them. We denote by $C_0$ the part of the service area not covered by a Wi-Fi wireless. Where from we have:

$$C_0 = C_1 - \bigcup_{j=2}^{m} C_j \tag{1}$$

As the users are equipped with devices of multiple accesses, they have the possibility to connect or disconnect from a network in the cells where networks overlap by choosing automatically the network which has the best bit rate. In our study, we suppose that the LTE network supplies two types of services: those Multicasts or Unicasts whose numbers of units of bandwidth are respectively $B_1^{mc}$ et $B_1^{uc}$. Besides, the number of units of bandwidth for every Wi-Fi wireless is $B_i$.

In the service area $C_1$, we suppose to have Q interferences sources distributed following a normal random distribution: $Q = \{I(q), q = 1, \ldots, Q\}$.

The selection technique is based on the bit rate then these interferences play an important role at the level of this strategy allowing to select a network. The different parameters of this heterogeneous networks system are given in the following Tables 1 and 2:

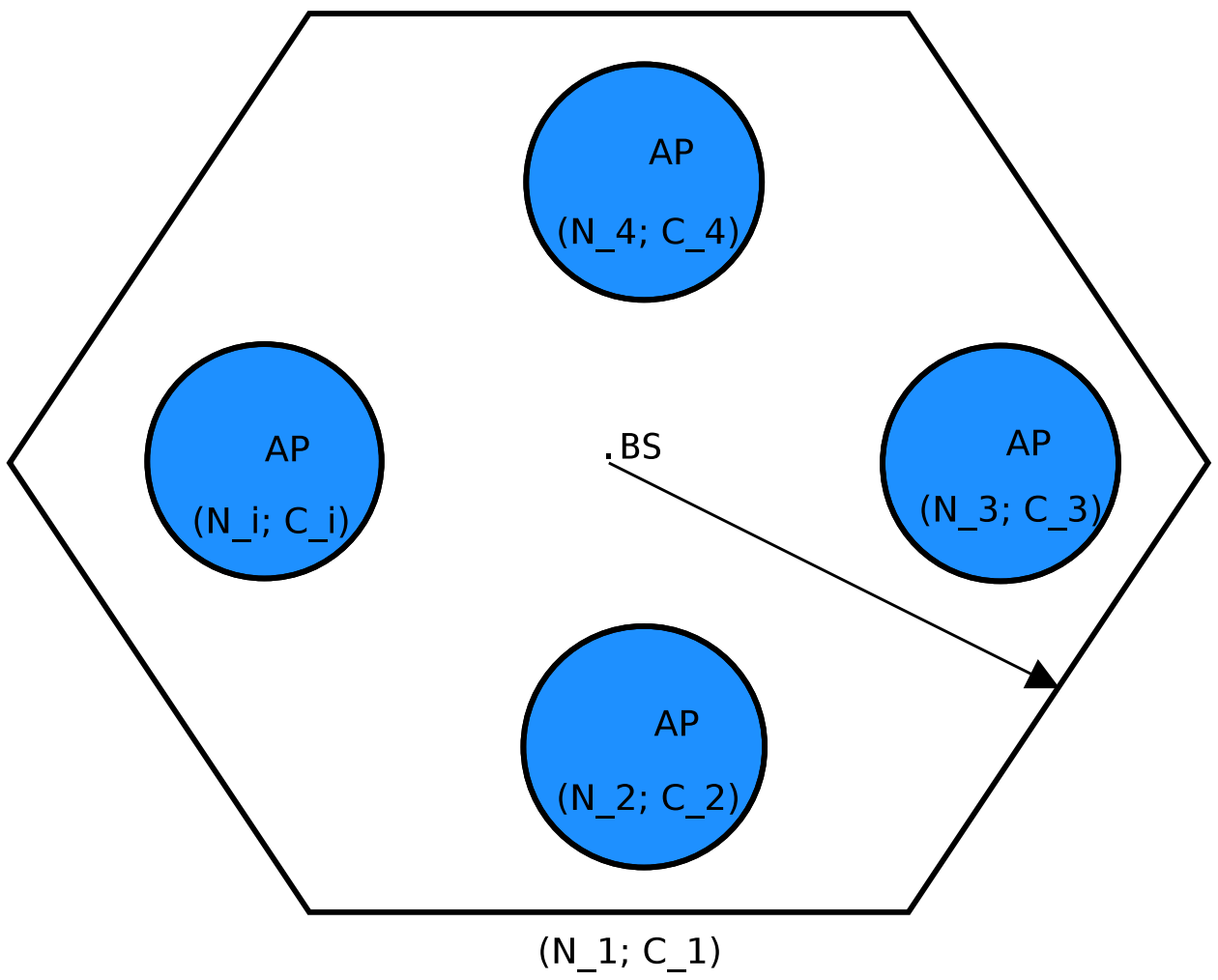

**Fig. 1** Model cluster service zone

**Table 1** LTE network parameters

| Parameters | LTE networks $N_1$ | |
|---|---|---|
| Interferences | Q sources | |
| Covered zone | $C_1$ | |
| Service quality | Low bandwidth | |
| Types of services | Multicast | Unicast |
| Units of bandwidth | $B_1^{mc}$ | $B_1^{uc}$ |

**Table 2** Wi-Fi wireless parameters

| Parameters | Wi-Fi wireless $N_{i=2,\ldots,m}$ |
|---|---|
| Interferences | Q sources |
| Covered zones | $C_{i=2,\ldots,m}$ |
| Service quality | Large bandwidth |
| Types of services | Unicast |
| Units of bandwidth | $B_i$ |

## Selection method

When a user is in a cell $C_i$ where both LTE mobile network and Wi-Fi wireless overlap then he has the possibility of connecting recently or by handover to the LTE network or Wi-Fi. If the LTE network has more free bandwidth units then user connects there otherwise he is blocked to connect to the Wi-Fi wireless as indicates Fig. 2. Besides, when a user wishes to connect recently in one sub-zone $C_i$ then his terminal makes the same technique to select that previously.

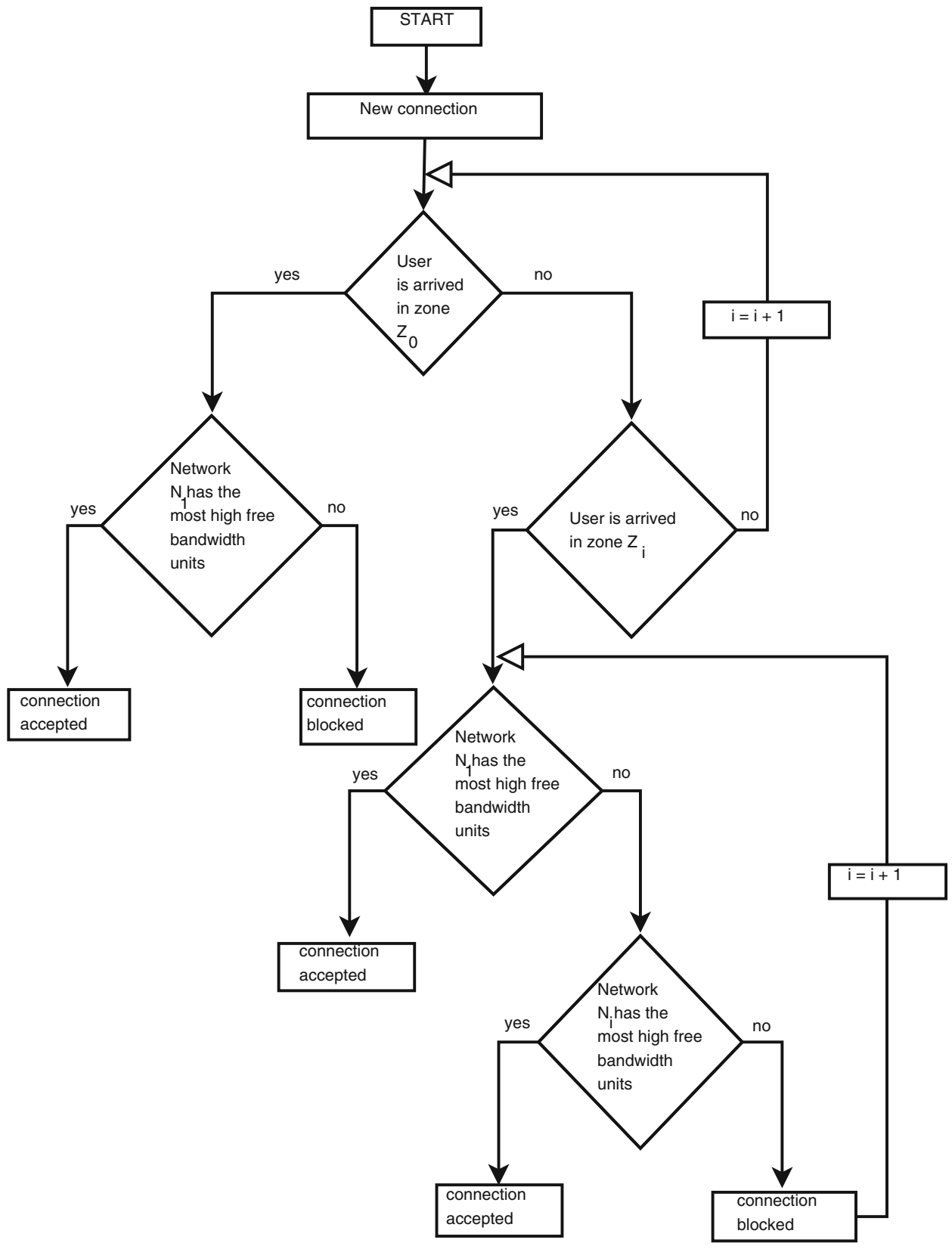

**Fig. 2** Selection technique

Beyond choices connections accepted and blocked, we can add the third choice which we would nickname: "the selection technique of a network is striggered". However, we supposed indirectly this choice without quoting it in particular and we do not consider necessary to add this choice in the plan of selection technique.

As the selection technique is based on the network which has free bandwidth units then the mobile terminal detects all the accesses points and all the base stations in its neighborhood then estimates automatically the bit rate of each of them. This one depends on the sub-carrier bandwidth and modulation.

## User mobility model

In the Random WayPoint (RWP) model, we suppose that a user moves in a convex space $C_1 \subset \mathbb{R}^2$ along a straight segment coming from any point towards another point. The various points which were occupied by the user are denoted by $P_i$. These points are uniformly distributed in $C_1$, $P_i \sim U(C_1)$. The transition of $P_{i-1}$ to $P_i$ corresponds to the $i_{th}$ straight segment, and the velocity of the user on this segment is defined as a random variable $v_i$, $v_i \sim v$. Particularly, the RWP model assures that variables $P_i$ and $v_i$ are independent. By basing itself of this notation, the process of the RWP for a user is defined by an infinite sequence of triplets,

$$\{(P_0, P_1, v_1), (P_1, P_2, v_2), \ldots, (P_{n-1}, P_n, v_n)\}$$

We can note that in the process of the RWP model, the consecutive straight segments of movement are independent because they share a point in common. However, let us note that this process is reversible in the time as far as, any path $P_0, P_1, \ldots, P_n$ is equivalent at time to the inverse path $P_n, P_{n-1}, \ldots, P_0$.

### Probability density

The probability density of finding a user situated at a distance $x$ of the convex cluster center $C_1$ and in a circular sub-cell of radius $r_i$ placed at a distance $d_i$ of the service area center is defined by [5]:

$$h(x) = (1 - x^2) \int_0^{\pi} \sqrt{1 - x^2 \cos(\Phi + \alpha - \beta)} \mathrm{d}\Phi \tag{2}$$

with

$$x = \sqrt{d_i^2 + 2 d_i r_i \cos\alpha + r_i^2}$$

$$\beta = \arctan(d_i + r_i \cos\alpha; r_i \sin\alpha)$$

### The arrival rate $\tau_a(C_i)$ of a user in a sub-cell $C_i$

By means Random WayPoint model and results obtained by [4], we have proved that the average arrived rate of a user in a sub-cell $C_i$ of radius $r_i$ and situated at a distance $d_i$ of the cluster center is established by equation:

$$\tau_a(Z_i) = \frac{2}{C_v}\int_0^{\pi}\int_0^{\pi} r_i.h(x)\sin\Phi \mathrm{d}\Phi \mathrm{d}\alpha \tag{3}$$

### Probability $\mathbb{P}(C_i)$ of finding users in a sub-cell $C_i$

The probability of finding users in a cell $C_i$ of radius $r_i$ situated at a distance $d_i$ of the service area center depends directly on the users mobility. Thus we are based on the Random WayPoint and works results of the authors [3] to show this probability by equation:

$$\mathbb{P}(C_i) = \int_0^{R}\int_0^{\pi}\int_0^{2\pi} r_i.h(x)\mathrm{d}\Phi \mathrm{d}\alpha \mathrm{d}r_i \tag{4}$$

## Average new access demand rate $\lambda_{C_i}^{C(k)}$ for a service

Let us denote by $\lambda_{C_1}^{C(k)}$ the average demand access rate for a service $k$ in a cell $C_1$. So the average demand access rate for a service $k$, $\lambda_{C_i}^{C(k)}$, being in one sub-cell $C_i$ is defined by formula:

$$\lambda_{C_i}^{C(k)} = \mathbb{P}(C_i)\cdot\lambda_{C_1}^{C(k)} \tag{5}$$

## Average demand rate of handovers

A connected user and then in mobility makes indirectly handover in the heterogeneous networks system. This handover is horizontal or vertical.

### Horizontal

Let $u_{C_0}^k$ the average user number having access to the service $k$ in the cell $C_0$. The average demands rate of horizontal handover $\tau_{C_0}^{H(k)}$ to the network $N_1$ for a service $k$ is given by relation 6 referring to the author (Jabban et al. [6]):

$$\tau_{C_1}^{H(k)} = u_{C_0}^k\cdot\eta_{C_0}^{C_1} \tag{6}$$

where $\eta_{C_0}^{C_1}$ is the users exit flow of the zone $C_0$ outside cell $C_1$ and is defined by:

$$\eta_{C_0}^{C_1} = \frac{\mathbb{P}(C_0)}{\Delta_{C(k)}}$$

with $\Delta_{C(k)}$ the average residence time of users in a zone $C_i$.

### Vertical

We denote by $u_{C_0}^k$ the average number of mobile users who have accessed to the service $k$ in the cell $C_0$. The average demands rate of vertical handover $\tau_{C_0}^{V(k)}$ to the network $N_i$ of users who have accessed to the service $k$ in the zone $C_0$ and moving towards the cell $C_i$ without finishing their connections is defined by Eq. 7 refer to the author (Jabban et al. [6]):

$$\tau_{C_i}^{V(k)} = u_{C_0}^k\cdot\eta_{C_0}^{C_i} \tag{7}$$

where $\eta_{C_0}^{C_i}$ is the users exit flow of the cell $C_0$ towards cell $C_i$ and is given by:

$$\eta_{C_0}^{C_i} = \frac{\mathbb{P}(C_0)}{\Delta_{C(k)}}$$

with $\Delta_{C(k)}$ the average residence time of the users in a zone $C_i$.

## Modeling approach based on a Markov chain

We have developed a Markov chain for modeling the dynamic fluctuations and define all the stages and states of the heterogeneous networks system.

Supposing that the set cells and services which are present in the service area, are respectively denoted $M$ and $S$ so the Markov chain size is given by : $s.(2m+1)$ with $\mid M\mid = m$ and $\mid S\mid = s$.

### Different stages and states of the system

When we take the system at the given moment then we define it as being a stage of dynamic change.

Besides, user connections and disconnections from a network give the various states of the system which define the states space by (see Table 3):

$$\mathcal{E} = \{(b_{1,1}^k; b_{1,2}^k;\ldots; b_{1,i}^k;\ldots; b_{1,m}^k;; b_2^k;\ldots; b_i^k;\ldots; b_m^k)\}$$

$$s.q/\begin{cases}\sum_{j=1}^{m}\sum_{k=1}^{s}(b_{1j}^k)\leq B_1^{uc}\\ \sum_{k=1}^{s}(b_i^k)\leq B_i\end{cases}$$

States are differentiated by the possible variation of the units of busy bandwidth in a given zone. For example when the system is in a given state, it changes state if a user connects recently or to make a handover either disconnects by freeing units of busy bandwidth.

**Table 3** Definition of system parameters

| Parameters | Definitions |
|---|---|
| $b_{1,1}^k$ | Number of units of busy bandwidth of LTE network to the service $k$ in the cell $C_0$ |
| $b_{1,i}^k$ | Number of units of busy bandwidth of LTE network to the service $k$ in the cell $C_i$ |
| $b_i^k$ | Number of units of busy bandwidth of LTE network to the service $k$ in the cell $C_i$ |
| $N_{\text{PRB}}^k$ | Number of resources blocks asked to supply a service $k$ by the LTE network in the cell $C_i$ |
| $\sum_{k=1}^{s} b_{1,1}^k = b_{1,1}$ | Number of units of busy bandwidth of LTE network in the cell $C_0$ |
| $\sum_{k=1}^{s} b_{1,i}^k = b_{1,i}$ | Number of units of busy bandwidth of LTE network in the cell $C_i$ |
| $\sum_{k=1}^{s} b_i^k = b_i$ | Number of units of busy bandwidth of Wifi wireless in the cell $C_i$ |

$$b_{11} \longrightarrow b_{11} + N_{\text{PRB}}^k$$

Where $b_{11}$ is the number of units of busy bandwidth in the zone $C_1$ by the network $N_1$ and the number of busy blocks resources by a user for a service $k$ in the zone $C_1$ to the network $N_1$.

- Stage 0: $E_0 = (b_{1,1}^k; \ldots; b_{1,i}^k; \ldots; b_{1,m}^k; b_2^k; \ldots; b_i^k; \ldots; b_m^k)$
- Stage 1: $E_1 = (b_{1,1}^k + \mu N_{prb}^k; b_{1,2}^k; \ldots; b_{1,m}^k; b_2^k; \ldots; b_m^k)$

$$\mu = \begin{cases} 1 & \text{if a user } \mathrm{u_0} \text{ connects to the network LTE in the cell } \mathrm{C_0} & \mathrm{State}(\mathrm{E_{1,1}}) \\ -1 & \text{if a user } \mathrm{u_0} \text{ disconnects of the LTE from the cell } \mathrm{C_0} & \mathrm{State}\ (\mathrm{E_{1,2}}) \\ 0 & \text{Otherwise} & \mathrm{State}(\mathrm{E_{1,3}}) \end{cases}$$

- Stage 2: $E_2 = (b_{1,1}^k; \ldots; b_{1,i}^k + \mu N_{prb}^k; \ldots; b_{1,m}^k; b_2^k; \ldots; b_m^k)$

$$\mu = \begin{cases} 1 & \text{if a user } \mathrm{u_0} \text{ connects to the network LTE in the cell } \mathrm{C_i} & \mathrm{State}(\mathrm{E_{2,1}}) \\ -1 & \text{if a user } \mathrm{u_0} \text{ disconnects of the network LTE from the cell } \mathrm{C_i} & \mathrm{State}(\mathrm{E_{2,2}}) \\ 0 & \text{Otherwise} & \mathrm{State}(\mathrm{E_{2,3}}) \end{cases}$$

- Stage 3: $E_3 = (b_{1,1}^k; \ldots; b_{1,m}^k; b_2^k; \ldots; b_i^k + \mu; \ldots; b_m^k)$

$$\mu = \begin{cases} 1 & \text{if a user } \mathrm{u_0} \text{ connects to the wireless Wi} - \text{Fi in the cell } \mathrm{C_i} & \mathrm{State}(\mathrm{E_{3,1}}) \\ -1 & \text{ifauser } \mathrm{u_0} \text{ disconnects of the wireless Wi} - \text{Fi from the cell } \mathrm{C_i} & \mathrm{State}(\mathrm{E_{3,1}}) \\ 0 & \text{Otherwise} & \mathrm{State}(\mathrm{E_{3,1}}) \end{cases}$$

- Stage 4: $E_4 = (b^k_{1,1} + \mu N^k_{PRB}; \ldots; b^k_{1,i} + \mu' N^k_{PRB}; \ldots; b^k_{1,m}; b^k_2; \ldots; b^k_m)$

$$(\mu, \mu') = \begin{cases} (1,-1) & \text{if a user } \mathrm{u_0} \text{ connects to the network LTE in the cell } \mathrm{C_0} \\ & \text{by disconnecting of network LTE from the cell } \mathrm{C_i} \qquad \mathrm{State}(\mathrm{E_{4,1}}) \\ (-1,1) & \text{if a user } \mathrm{u_0} \text{ disconnects of the network LTE from the cell } \mathrm{C_0} \\ & \text{by connecting to the network LTE in the cell } \mathrm{C_i} \qquad \mathrm{State}(\mathrm{E_{4,2}}) \\ (0,0) & \text{Otherwise} \qquad \mathrm{State}(\mathrm{E_{4,3}}) \end{cases}$$

- Stage 5: $E_4 = (b^k_{1,1} + \mu N^k_{PRB}; b^k_{1,2}; \ldots; b^k_{1,m}; b^k_2; \ldots; b^k_i + \mu'; \ldots; b^k_m)$

$$(\mu, \mu') = \begin{cases} (1,-1) & \text{if a user } \mathrm{u_0} \text{ connects to network LTE in the cell } \mathrm{C_0}\text{by} \\ & \text{disconnecting of the wireless Wi} - \text{Fi from the cell } \mathrm{C_i} \qquad \mathrm{State}(\mathrm{E_{5,1}}) \\ (-1,1) & \text{if a user } \mathrm{u_0} \text{ disconnects of the network LTE from the cell } \mathrm{C_0} \\ & \text{by connecting to the wireless Wi} - \text{Fi in the cell } \mathrm{C_i} \qquad \mathrm{State}(\mathrm{E_{5,2}}) \\ (0,0) & \text{Otherwise} \qquad \mathrm{State}(\mathrm{E_{5,3}}) \end{cases}$$

- Stage 6: $E_5 = (b^k_{1,1}; b^k_{1,2}; \ldots; b^k_{1,i} + \mu N^k_{PRB}; \ldots; b^k_{1,m}; b^k_2; \ldots; b^k_i + \mu'; \ldots; b^k_m)$

$$(\mu, \mu') = \begin{cases} (1,-1) & \text{if a user } \mathrm{u_0} \text{ connects to the network LTE in the cell } \mathrm{C_i} \text{ by} \\ & \text{disconnecting of the wireless Wi} - \text{Fi from the cell } \mathrm{C_i} \qquad \mathrm{State}(\mathrm{E_{6,1}}) \\ (-1,1) & \text{if a user } \mathrm{u_0} \text{ disconnects of the network LTE from the cell } \mathrm{C_i} \\ & \text{by connecting to the wireless Wi} - \text{Fi in the cell } \mathrm{C_i} \qquad \mathrm{State}(\mathrm{E_{6,2}}) \\ (0,0) & \text{Otherwise} \qquad \mathrm{State}(\mathrm{E_{6,3}}) \end{cases}$$

## Transition rate

**Proposition 1** Let us consider the heterogeneous system of networks in a stage $(E_p)_{1 \le p \le 6}$. The transition rate towards the state $E_{p,t}$ of the stage $E_p$ is established by:

(1) In the zone $C_0$ of the cluster:

$$\tau_{(E_0 \rightleftarrows E_{1,1})} = \tau_{1,1} = (\lambda^{C(k)}_{C_0} + \lambda^{H(k)}_{N_1}) \cdot \left( \frac{b^k_{11}}{N^k_{PRB}} + 1 \right) \cdot \left( \frac{1}{\Delta_{C(k)}} + \eta^{\overline{C_1}}_{C_1} \right) \tag{8}$$

(2) In the cell $C_i$ of the cluster:

$$\tau_{(E_1 \rightleftarrows E_{2,1})} = \tau_{2,1} = (\lambda_{C_i}^{c(k)} + \lambda_{C_i}^{c(k)} \cdot \mathbb{P}(N_1 \to N_i)) \cdot \left(\frac{b_{1i}^k}{N_{PRB}^k} + 1\right) \cdot \frac{1}{\Delta_{c(k)}} \quad (9)$$

(3) From the zone $C_0$ to the cell $C_i$ of cluster:

- By horizontal handover:

$$\tau_{(E_3 \rightleftarrows E_{4,1})} = \tau_{4,1} = \left(\frac{b_{11}^k}{N_{PRB}^k} + 1\right) \cdot \left(\frac{b_{1i}^k}{N_{PRB}^k}\right) \cdot \tau_{(C_0 \leftarrow C_i)}^{H(k)} \quad (10)$$

- By vertical handover:

$$\tau_{(E_4 \rightleftarrows E_{5,1})} = \tau_{5,1} = (b_i^k) \cdot \left(\frac{b_{11}^k}{N_{PRB}^k} + 1\right) \cdot \tau_{(C_0 \leftarrow C_i)}^{V(k)} \quad (11)$$

*Proof*

(1) When the number of units of busy bandwidth in one under zone $C_1$ of the network $N_1$ (LTE) crosses of $b_(11)$ to $b_{11} + N_{\text{PRB}}^k$ then the variation rate of the units of busy bandwidth is defined by:

$$\frac{b_{11} + N_{\text{PRB}}^k}{N_{\text{PRB}}^k} = \frac{b_{11}}{N_{\text{PRB}}^k} + 1$$

If there is variation of units of busy bandwidth, it is because a user who is recently connected in it in sub-zone $C_1$ with a access demand rate $\lambda_{C_0}^{C(k)}$ or to make a horizontal handover $\lambda_{N_1}^{H(k)}$ and the residence time in this sub-zone $C_1$ is $\Delta_{C(k)}$ or the time of passage in this cell $C_0$ towards $C_1$. The residence time is inversely proportional to the transition rate. The transition rate of the state $E_0$ to the state $E_1$ is equal to the product of these various rates calculated previously. So the found rate is:

$$(\lambda_{C_0}^{C(k)} + \lambda_{N_1}^{H(k)}) \cdot \left(\frac{b_{11}^k}{N_{\text{PRB}}^k} + 1\right) \cdot \left(\frac{1}{\Delta_{C(k)}} + \eta_{C_0}^{C_1}\right)$$

(2) If a user $u$ connects to the network $N_1$ (LTE) in a sub-zone $C_i$ then the number of bandwidth units occupied in this sub-zone passes to $b_{1i}$ $b_{1i} + N_{PRB}^k$. So bandwidth units rate variation occupied in this sub-zone is calculated by:

$$\frac{b_{1i} + N_{\text{PRB}}^k}{N_{\text{PRB}}^k} = \frac{b_{1i}}{N_{\text{PRB}}^k} + 1$$

This variation of the bandwidth units occupation rate is due to the fact that when a user is recently connected in this sub-zone $C_i$ with an access request rate $\lambda_{C_i}^{c(k)}$ rr disconnects from the network $N_i$ by connecting to the network $N_1$ with a probability $\mathbb{P}(N_i \to N_1)$ applied to an access request rate $\lambda_{C_i}^{c(k)}$ in this sub-zone. The transition rate to the state $E_{2,1}$ of the stage $E_2$ being inversely proportional of stay time $\Delta_{c(k)}$ in this su-zone $C_i$ then it's found by making the product of the various rates calculated previously. So the rate $\tau_{(E_1 \rightleftarrows E_{2,1})}$ found is equal to:

$$(\lambda_{C_i}^{c(k)} + \lambda_{C_i}^{c(k)} \cdot \mathbb{P}(N_1 \to N_i)) \cdot \left(\frac{b_{1i}^k}{N_{PRB}^k} + 1\right) \cdot \frac{1}{\Delta_{c(k)}}$$

(3) Besides, if a user $u$ connects to the network $N_1$ (LTE) in sub-zone $C_0$ by disconnecting from the same network $N_1$ but in a zone $C_i$ then the number of bandwidth units occupied in this sub-zone $C_0$ of the network $N_1$ (LTE) passes from $b_{11}$ to $b_{11} + N_{\text{PRB}}^k$ and $b_{1i}$ to $b_{1,i}^k - N_{\text{PRB}}^k$ in sub-zone $C_i$. So the variation rate of the bandwidth units occupied in zones $C_0$ and $C_i$ are respectively defined by:

$$\frac{b_{11} + N_{\text{PRB}}^k}{N_{\text{PRB}}^k} = \frac{b_{11}}{N_{\text{PRB}}^k} + 1$$

and

$$\frac{b_{1i}^k}{N_{\text{PRB}}^k}$$

As the user makes a horizontal handover(vertical resp.) of the zone $C_i$ towards $C_1$ then the rate of transition is directly proportional at the horizontal handover rate (vertical resp.) $\tau_{(C_0 \leftarrow C_i)}^{H(k)}$(resp. $\tau_{(C_0 \leftarrow C_i)}^{V(k)}$). So the transition rate of the state $E_(4,1)$ in the stage $E_3$ is equal to the product of these rates calculated previously. Thus the found rate is:

$$\left(\frac{b_{11}^k}{N_{\text{PRB}}^k} + 1\right) \cdot \left(\frac{b_{1i}^k}{N_{\text{PRB}}^k}\right) \cdot \tau_{(C_0 \leftarrow C_i)}^{H(k)}$$

Respectively:

$$((b_i^k) \cdot \left(\frac{b_{11}^k}{N_{\text{PRB}}^k} + 1\right) \cdot \tau_{(C_0 \leftarrow C_i)}^{V(k)})$$

□

**The various transitions:**

- $\tau_{(E_0 \rightleftarrows E_{1,1})} = (\lambda_{C_0}^{C(k)} + \lambda_{N_1}^{H(k)}) \cdot \left(\frac{b_{11}^k}{N_{\text{PRB}}^k} + 1\right) \cdot \left(\frac{1}{\Delta_{C(k)}} + \eta_{C_0}^{C_1}\right)$ $s.q/ \sum_{j=1}^{m} (b_{1j} + N_{\text{PRB}}^k) \leq B_1^{\text{uc}}$
- $\tau_{(E_0 \rightleftarrows E_{1,2})} = (\lambda_{C_0}^{C(k)} + \lambda_{N_1}^{H(k)}) \cdot \left(\frac{b_{11}^k}{N_{\text{PRB}}^k}\right) \cdot \left(\frac{1}{\Delta_{C(k)}} + \eta_{C_0}^{C_1}\right)$ $s.q/ \quad b_{11}^k \geq N_{\text{PRB}}^k$

- $\tau_{(E_1 \rightleftarrows E_{2,1})} = (\lambda_{C_i}^{C(k)} + \lambda_{C_i}^{C(k)} \cdot \mathbb{P}(N_1 > N_i)) \cdot \left(\frac{b_{1i}^k}{N_{\text{PRB}}^k} + 1\right) \cdot \frac{1}{\Delta_{C(k)}} \quad s.q/ \begin{cases} b_i = B_i \\ \sum_{j=1}^m (b_{1j} + N_{\text{PRB}}^k) \leq B_1^{\text{uc}} \end{cases}$
- $\tau_{(E_1 \rightleftarrows E_{2,2})} = (\lambda_{C_i}^{C(k)} + \lambda_{C_i}^{C(k)} \cdot \mathbb{P}(N_1 > N_i)) \cdot \frac{b_{1i}^k}{N_{\text{PRB}}^k} \cdot \frac{1}{\Delta_{C(k)}} \quad s.q/ \begin{cases} b_i = B_i \\ b_{1i}^k \geq N_{\text{PRB}}^k \end{cases}$
- $\tau_{(E_2 \rightleftarrows E_{3,1})} = (\lambda_{C_i}^{C(k)} + \lambda_{C_i}^{C(k)} \cdot \mathbb{P}(N_1 > N_i)) \cdot (b_i^k + 1) \cdot \frac{1}{\Delta_{C(k)}} \quad s.q/ \begin{cases} b_i \leq B_i \\ \sum_{j=1}^m (b_{1j} + N_{\text{PRB}}^k) = B_1^{\text{uc}} \end{cases}$
- $\tau_{(E_2 \rightleftarrows E_{3,2})} = (\lambda_{C_i}^{C(k)} + \lambda_{C_i}^{C(k)} \cdot \mathbb{P}(N_1 > N_i)) \cdot (b_i^k) \cdot \frac{1}{\Delta_{C(k)}} \quad s.q/ \begin{cases} b_i \geq 1 \\ \sum_{j=1}^m (b_{1j} + N_{\text{PRB}}^k) = B_1^{\text{uc}} \end{cases}$
- $\tau_{(E_3 \rightleftarrows E_{4,1})} = \left(\frac{b_{11}^k}{N_{\text{PRB}}^k} + 1\right) \cdot \left(\frac{b_{1i}^k}{N_{\text{PRB}}^k}\right) \cdot \tau_{(C > C_i)}^{H(k)} \quad s.q/ \begin{cases} b_i = B_i \\ \sum_{j=1}^m (b_{1j} + N_{\text{PRB}}^k) \leq B_1^{\text{uc}} \\ b_{1i}^k \geq N_{\text{PRB}}^k \end{cases}$
- $\tau_{(E_3 \rightleftarrows E_{4,2})} = \left(\frac{b_{11}^k}{N_{\text{PRB}}^k}\right) \cdot \left(\frac{b_{1i}^k}{N_{\text{PRB}}^k} + 1\right) \cdot \tau_{(C_i > C_0)}^{H(k)} \quad s.q/ \begin{cases} b_i = B_i \\ \sum_{j=1}^m (b_{1j} + N_{\text{PRB}}^k) \leq B_1^{\text{uc}} \\ b_{11}^k \geq N_{\text{PRB}}^k \end{cases}$
- $\tau_{(E_4 \rightleftarrows E_{5,1})} = (b_i^k) \cdot \left(\frac{b_{11}^k}{N_{\text{PRB}}^k} + 1\right) \cdot \tau_{(C_i > C_0)}^{V(k)} \quad s.q/ \begin{cases} b_i \geq 1 \\ \sum_{j=1}^m (b_{1j} + N_{\text{PRB}}^k) \leq B_1^{\text{uc}} \end{cases}$
- $\tau_{(E_4 \rightleftarrows E_{5,2})} = (b_i^k + 1) \cdot \left(\frac{b_{11}^k}{N_{\text{PRB}}^k}\right) \cdot \tau_{(C_0 > C_i)}^{V(k)} \quad s.q/ \begin{cases} b_i \leq B_i \\ b_{1i}^k \geq N_{PRB}^k \end{cases}$
- $\tau_{(E_5 \rightleftarrows E_{6,1})} = \left(\frac{b_{1i}^k}{N_{\text{PRB}}^k} + 1\right) \cdot (b_i^k) \quad s.q/ \begin{cases} b_i \geq 1 \\ \sum_{j=1}^m (b_{1j} + N_{\text{PRB}}^k) \leq B_1^{\text{uc}} \end{cases}$
- $\tau_{(E_5 \rightleftarrows E_{6,2})} = \left(\frac{b_{1i}^k}{N_{\text{PRB}}^k}\right) \cdot (b_i^k + 1) \quad s.q/ \begin{cases} b_{1i}^k \geq N_{\text{PRB}}^k \\ b_i \leq B_i \end{cases}$

## Evaluations of optimal performances

We estimate the selection strategy performances of a network based on the bit rate related to the parameters such as the blocking probability and the connections quality.

### Average bit rate of the system in a sub-zone $C_i$

The bit rate received in a zone $Z_i$ from the network $N_1$ is in function of the number of present units of bandwidth in networks $N_1$ and $N_i$. By denoting $D_1^{\text{avg}}(E)$ the average bit rate value received from the network and $\mathbb{P}(E)$ the probability of system balance state so the total average bit rate value received in a zone $Z_i$ from the network $N_1$ is given by:

$$D_{N_1}^{\text{tot}} = \sum_{k=1}^{s} (\lambda_{C_i}^{c(k)} + \mathbb{P}(C_i > C_1)) \cdot \mathbb{P}(E) \cdot D_1^{\text{avg}}(E) \tag{12}$$

$$s.q/ \quad \sum_{j=1}^{m} (b_{1j} + N_{\text{PRB}}^k) \leq B_1^{\text{uc}}$$

with:

$$D_1^{\text{avg}}(E) = D_1^{\text{avg}} \cdot \left(1 - \Lambda \cdot \sqrt{\frac{\sum_{j=1}^m (b_{1j} + b_j)}{B_1^{\text{uc}} + B_i}}\right)$$

where the average instantaneous bit rate $D_1^{\text{avg}}(E)$ is defined by the product of a sub-carrier bandwidth and modulation(numbers of modulated sub-carriers):

$$D_1^{\text{avg}} = B_{\text{sp}} \times N_{\text{sub}} \tag{13}$$

where $N_{\text{sub}} = K \times B \times E_i \times (1 - \text{BLER}_i)$ such as:

- K is the frequencies number;
- B is the numbers of symbols per second;
- $E_i$ is the modulation efficiency;
- $\text{BLER}_i$ is the Block error rate in a sub-cell $C_i$;

As result we have:

$$\overline{D}_{C_{1i}} = \frac{D_{N_1}^{\text{tot}}}{\eta_{C_i}^{c(k)}} \tag{14}$$

### Blocking probability et connections losses in sub-cell $C_i$

We have calculated the mean blocking probability of connections in a sub-zone $C_i$ in function of an equilibrium sate probability $\mathbb{P}(E)$ of the system. Indeed, we added the system states probability where the numbers of units of busy bandwidth is higher than these available in the network $N_1$ by formula:

$$\mathbb{P}_{N_1}^B = \sum_{k=1}^{s} (\lambda_{C_i}^{C(k)} + \mathbb{P}(C_i > C_1)) \cdot \mathbb{P}(E) \cdot \mathbb{P}_1^B(E) \tag{15}$$

$$s.q/ \quad \sum_{j=1}^{m} (b_{1j} + N_{\text{PRB}}^k) > B_1^{\text{uc}}$$

with

$$\mathbb{P}_1^B(E) = \mathbb{P}_1^B \cdot \left(1 - \Theta \cdot \sqrt{\frac{\sum_{j=1}^m (b_{1j} + b_j)}{B_1^{\text{uc}} + B_i}}\right)$$

**Table 4** LTE network parameters test

| Test | Value |
|---|---|
| Modulation | 16 QAM |
| Symbols | 6 |
| Efficiency | 1.4766 |
| Sub-carriers number | 72 |
| Bandwidth | 1.4 MHz |

**Table 5** System network parameters test

| Parameters test | Value |
|---|---|
| Average access rate to a service | 70 % |
| Average handover rate | 60 % |
| Average state balance rate | 80 % |
| Bandwidth units in LTE | 60 |
| Noise power | −174 dBm/Hz |
| Signal power | 400 dBm |
| Services number | Two services unicast |
| Cell radius | 600 m $Z_1$, 200 m $Z_2$ |
| Distance between area centers | 300 m |

$$\mathbb{P}_1^B = \frac{\frac{\rho_{C_i}^s}{s!}}{\sum_{k=1}^{s} \frac{\rho_{C_i}^k}{k!}} \qquad (16)$$

where $s$ is the number of available services in the cluster and

$$\rho_{C_i} = \sum_{k=1}^{s} \left( \frac{\lambda_{C_i}^{c(k)}}{\lambda_{C_1}^{c}} \right)$$

is the probability that a user is blocked in the sub-zone $C_i$.

## Numerical tests

To test the theoretical results that we obtained, we simplified our field of study in a service cell $C_1$ covered by the network $N_1$ (LTE) in which we implanted a wireless $N_2$ (Wi-Fi) in sub-cell $C_2$ of $C_1$. By means of the simulator NS3 [9] and working with the parameters below we managed to obtain satisfactory results as show by the obtained curves.

We have supposed that the data used for the test are the ones relative to the characteristics of the network LTE. So, we chose the modulation 16QAM of the network LTE the number of symbols which is six (06), Efficiency is equal 1.4766, sub-carriers number is 72 and as the bandwidth of the network LTE varies between 1.4 and 20 MHz then we have worked with the minimal value (1.4MHz) as indicated in Table 4; refer to Jabban [8].

However, the parameters chosen for the heterogeneous mobile and wireless system of networks are supposed to allow us to have satisfactory results for the numerical results. These parameters are given in Table 5.

The main parameters of both available networks are illustrated in the Table 5. We suppose that the coverage radius of networks $N_1$ and $N_2$ are respectively equal to 600 M and 200 m. We also suppose that the power transmitted by $N_1$ is equal to 400 dBM. The noise power is supposed equal to −174 dBm/Hm for the network LTE. We also suppose that the average access requests rates for a service, handover and the system balance state are, respectively, 70, 60 and 80 %. The number of units of bandwidth is supposed equal to 60. Finally the number of unicasts services is fixed to 2.

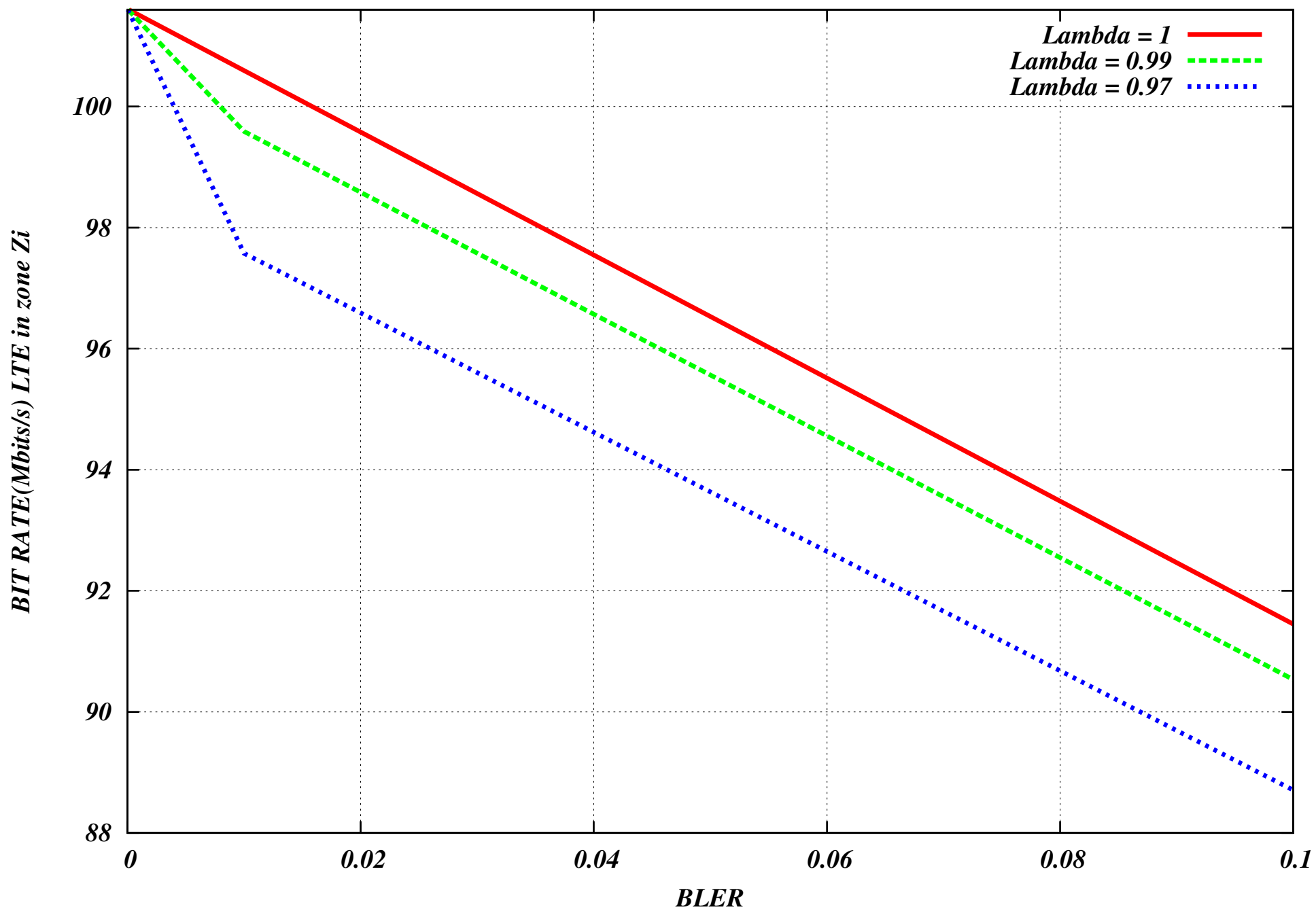

**Fig. 3** Average bit rate of network $N_1$ in the cell $C_2$ in function of the BLER

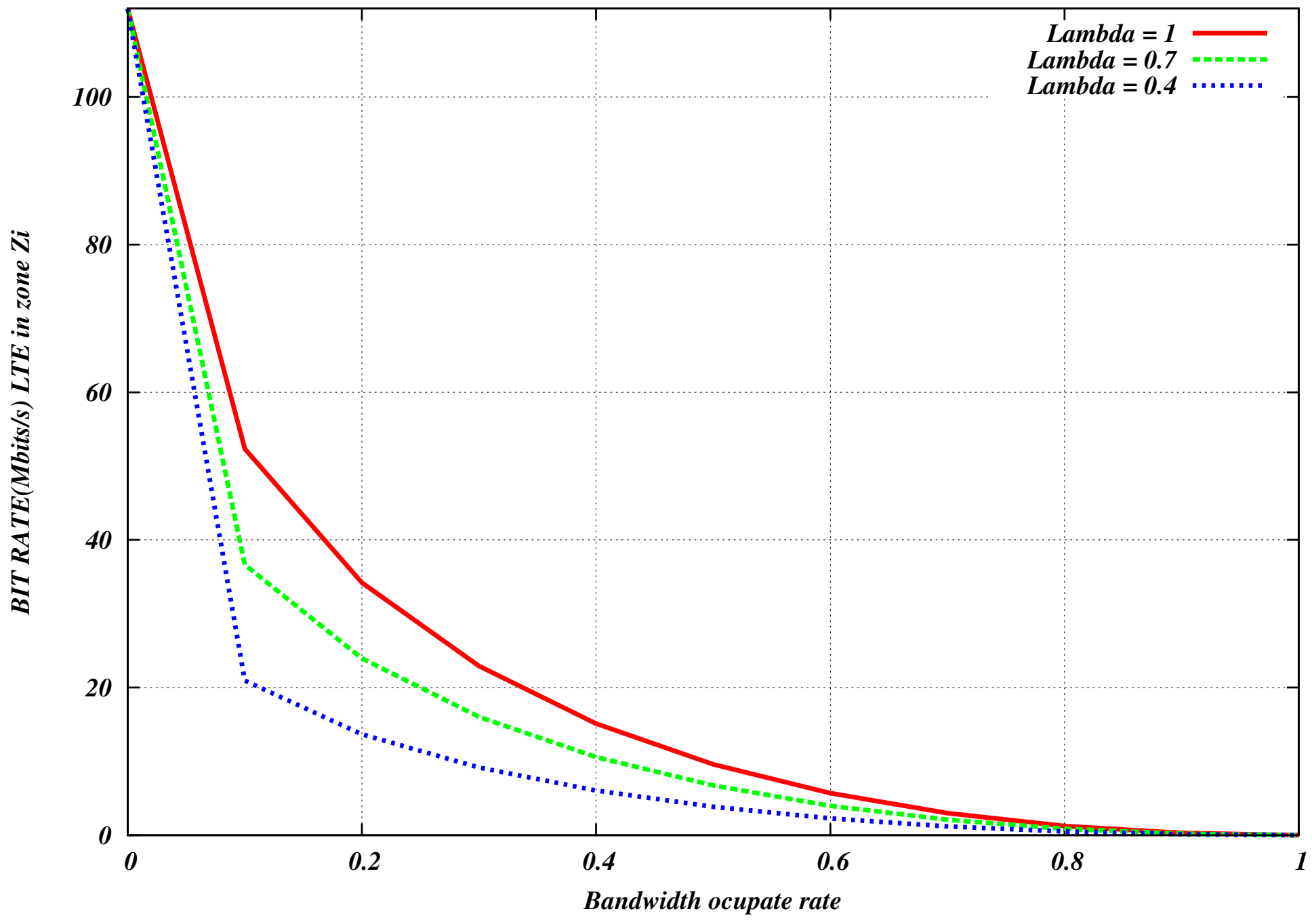

**Fig. 4** Average bit rate of the network $N_1$ in the zone $C_2$ in function of the busy bandwidth rate

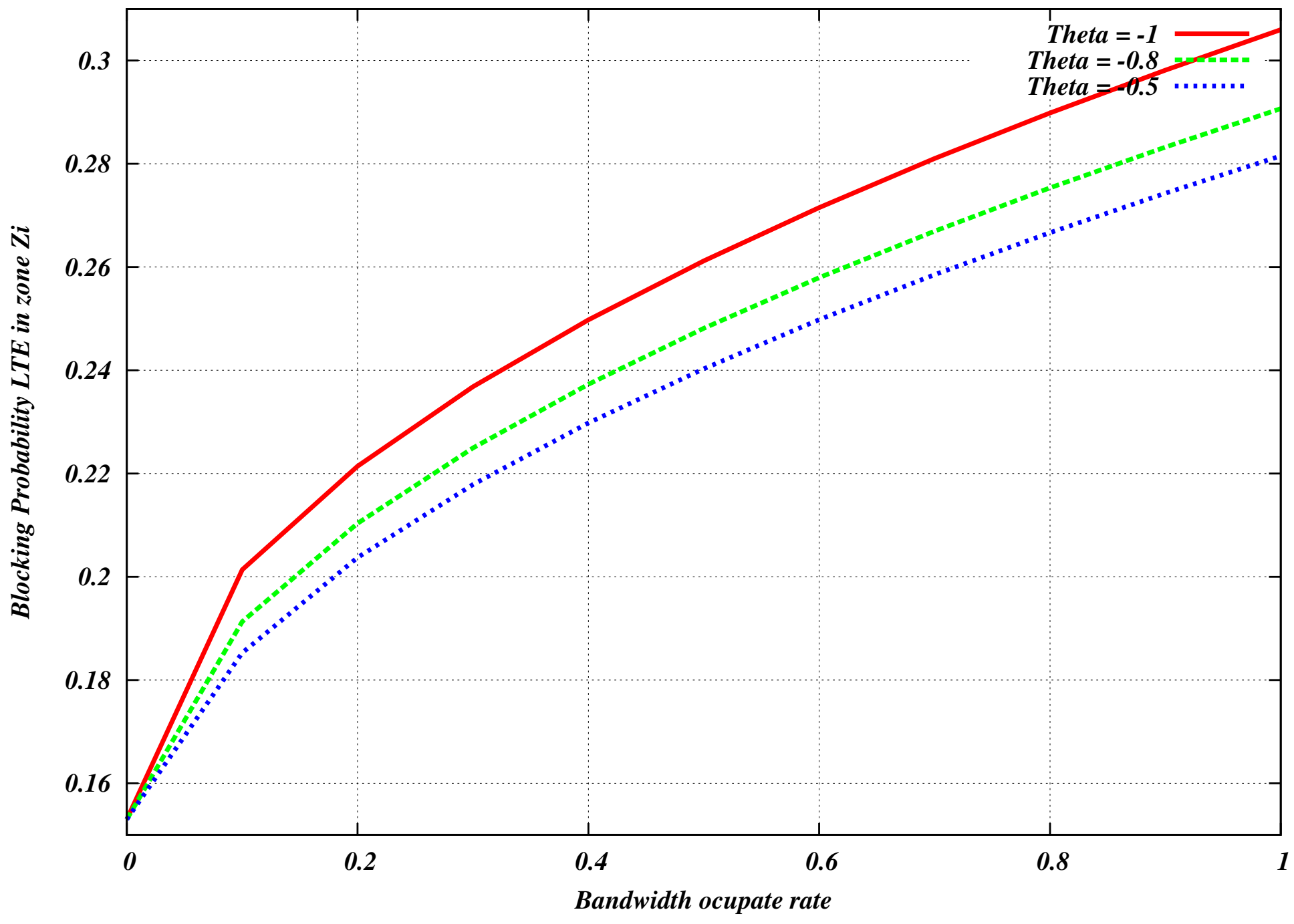

**Fig. 5** Blocking probability in the zone $C_2$ in function of the busy bandwidth rate

We have calculated the average bit rate value $D_{N_1}^{tot}$ received by a user $u_0$ from the network $N_1$ in the cell $C_2$. The selection strategy is based on the most high bit rate, in a sub-cell $C_2$, the user chooses the unicasts services of the networks $N_1$ or $N_2$ according to the selection technique established. Through the obtained results, we have noticed that a small modification of the bit rate occurs due to the network saturation. This modification is represented by a parameter $\Lambda$ which indicates a very high sensitivity of the bit rate because of the network congestion. For a low modification of the sensitivity factor $\Lambda$ for example from 1 to 0.99 we obtained a net fall of the bit rate as illustrate by Fig. 2 in function of the BLER in the sub-cell $C_2$. This factor also informs about the maximal decrease of the bit rate because of the network congestion.

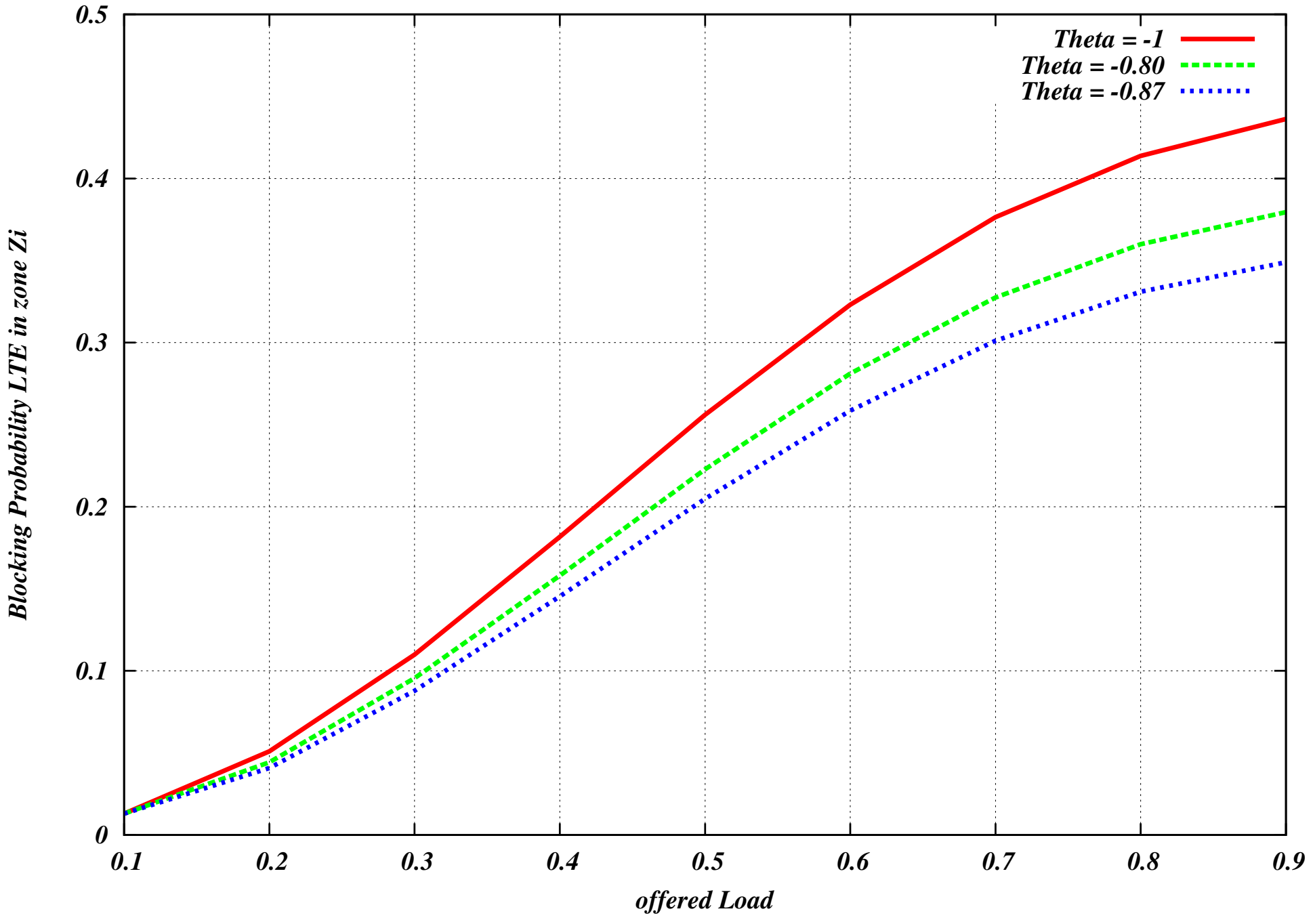

**Fig. 6** Blocking probability in the zone $C_2$ in function of the offered load rate

However the change of the factor $\Lambda$ acts less on the bit rate when it is estimated in function of the occupation bandwidth rate by fixing the BLER as indicated by Fig. 3; BLER is fixed to 50 %.

Besides, we have estimated the network performances related to the average blocking probability and connections losses to the services in the sub-cell $C_2$. The obtained results depend on a sensitivity factor $\Theta$ as represented by Figs. 4 and 5. For sensitivity parameters $\Theta \in \{1; 0.8; 0.5\}$ we analyzed the blocking probability in the sub-cell $C_2$ in function of the busy bandwidth rate (Fig. 4) then to the offered load by the traffic (Fig. 5). Indeed, seen the satisfactory obtained results, we realize that the blocking probability do not pass 40 % when they are determined with the occupation bandwidth rate. They do not reach either the level 50 % when they are estimated in function of the offered load by the traffic whatever is the given sensitivity factor (Fig. 6).

## Conclusion

At the end of our analysis on the system integration performances of new generation wireless and mobile networks, we found a factor which remains very sensitive to the variations of the bit rate received in a sub-cell $C_i$ when it is calculated in function of the BLER in this sub-zone. Besides, this sensitivity factor remains so determining for the blocking probability theory in a sub-zone $C_i$ by means of the busy bandwidth rate or the offered load traffic rate. The satisfactory results obtained on the system performances of wireless and mobile networks such the LTE and the Wi-Fi based on the bit rate allowed us to discover the dynamic fluctuations system. The parameters related to the bit rate such as the blocking probability are estimated with lower rates the bar 40 %.

In our future works, we intend to calculate the same sensitivity factor when we consider the performances of the system related to the SINR. We planned also to take into account the number of users having consumed these numbers of busy bandwidth units. This will allow us to encircle better the congestion and dis-congestion rates of the heterogeneous networks.

**Acknowledgments** We are solemnly anxious to thank all the members of research team of mathematical decision of Cheikh Anta Diop University for their collaboration, their critic and their support to realize this project. We also thank all the mathematics teachers of the department of mathematics and computing at Cheikh Anta Diop university of Dakar for the realization of these scientific works.